\documentclass[twocolumn]{aastex63}

\usepackage{amssymb}
\usepackage{amsmath}
\usepackage{amsfonts}
\usepackage{graphicx}
\usepackage{color}
\usepackage{hyperref}

\def\lsim{~\rlap{$<$}{\lower 1.0ex\hbox{$\sim$}}}
\def\bsim{~\rlap{$>$}{\lower 1.0ex\hbox{$\sim$}}}

\def\apjl{Astrophys.\ J.\ Lett.}

\def\aap{Astron.\ Astrophys.}

\def\apjs{Astrophys.\ J. Supp.}

\def\mathbi#1{\textbf{\em #1}}

\def\kvh{\mathrm{\hat{\bf{k}}}}

\def\pvh{\hat{\boldsymbol{\varphi}}}

\def\rvh{\mathrm{\hat{\bf{r}}}}
\def\xvh{\mathrm{\hat{\bf{x}}}}
\def\yvh{\mathrm{\hat{\bf{y}}}}
\def\zvh{\mathrm{\hat{\bf{z}}}}

\def\ve{\mathbi{e}}
\def\vk{\mathbi{k}}

\def\vr{\mathbi{r}}

\def\vu{\mathbi{u}}
\def\vv{\mathbi{v}}

\def\vx{\mathbi{x}}

\def\dirac{\delta_D}

\def\fdf{\mathbi{F}_\text{DF}}
\def\mach{\mathcal{M}}

\allowdisplaybreaks

\usepackage[normalem]{ulem}

\begin{document}

\title{Analytic solution to the dynamical friction acting on circularly moving perturbers}

\author{Vincent Desjacques}
\affiliation{Physics Department and Asher Space Science Institute, Technion -- Israel Institute of Technology, Haifa 3200003, Israel}
\author{Adi Nusser}
\affiliation{Physics Department and Asher Space Science Institute, Technion -- Israel Institute of Technology, Haifa 3200003, Israel}
\author{Robin B\" uhler}
\affiliation{Physics Department and Asher Space Science Institute, Technion -- Israel Institute of Technology, Haifa 3200003, Israel}

\correspondingauthor{Vincent Desjacques}
\email{dvince@physics.technion.ac.il}

\date{\today}

\begin{abstract}

  We present  an analytic approach to  the dynamical friction (DF) acting on a circularly moving point mass perturber in a gaseous medium.
  We demonstrate that, when the perturber is turned on at $t=0$, steady-state (infinite time perturbation) is achieved after exactly one sound-crossing time. 
  At low Mach number $\mach~\ll~1$, the circular-motion steady-state DF converges to the linear-motion, finite time perturbation expression. 
  The analytic results describe both the  radial and tangential forces on the perturbers caused by the backreaction of the wake propagating in the medium. The radial force is directed inward, toward the motion centre, and is dominant at large Mach numbers. For subsonic motion, this component is negligible. For moderate and low Mach numbers, the tangential force is stronger and opposes the motion of the perturber.
  The analytic solution to the circular-orbit DF suffers from a logarithmic divergence in the supersonic regime. This divergence appears at short distances from the perturber solely (unlike the linear motion result which is also divergent at large distances) and can be encoded in a maximum multipole. This is helpful to assess the resolution dependence of numerical simulations implementing DF at the level of Li\'enard-Wiechert potentials.
  We also show how our approach can be generalised to calculate the DF acting on a compact circular binary.

\end{abstract}

\section{Background}

The gravitational force field of a massive object (perturber) moving in a discrete or continuous  medium induces a density fluctuation (wake)  in that medium. 
Dynamical friction (DF) is the gravitational  force exerted on the perturber as a result of the induced density field.
The pioneering study of \cite{chandrasekhar:1943} considered a perturber linearly moving in a collisionless medium, whereas \cite{bondi/hoyle:1944,dokuchaev:1964,ruderman/spiegel:1971,rephaeli/salpeter:1980,just/kegel:1990,ostriker:1999} focused on a gaseous medium. \cite{barausse:2007}
and \cite{namouni:2010} extended these analytical results to linear relativistic motion and constant linear acceleration, respectively. Other studies include, e.g., DF in fuzzy dark matter backgrounds \citep{hui/etal:2016,lancaster/etal:2020,annulli/etal:2020} or superfluids \citep{berezhiani/eal:2019}.

For   a perturber in a linear motion at all times (steady state at any time) in a gaseous medium, DF is absent for Mach numbers $\mach<1$, while it exhibits a logarithmic divergence for $\mach>1$ as in Chandrasekhar's formula for a collisionless medium. For the finite time perturbation turned on at some initial time $t=0$, DF scales like $\mach^3$ for subsonic motions and still features a Coulomb logarithm for $\mach>1$ \citep{just/kegel:1990,ostriker:1999}. Furthermore, linear steady-state motion is never achieved.

Frictional effects have been investigated in the case of a perturber circularly moving in a collisionless stellar system \citep[see, e.g.,][]{tremaine/weinberg:1984,kaur/sridhar:2018,banik/vandenbosch:2021}. 
However, these authors did not consider the radial force on the perturber.
Furthermore, no analytical solution has been derived thus far in the gaseous case, in spite of its potentially wide astrophysical implications.  A numerical investigation of DF for a   circular motion in a gaseous medium  is performed by \cite{sanchez/brandenburg:2001,kim/etal:2007,kim/etal:2008}. 
Ostriker’s formula \citep{ostriker:1999} for linear motion is found to furnish a reasonable estimate of the tangential DF (drag) provided that the outer cutoff radius in the Coulomb logarithm is set to the orbital diameter. However, a small region around the perturber had to be excised in the supersonic case to regularize a small-scale divergence. This suggested that the circular DF also features a Coulomb logarithm. 

In this paper, we derive an analytical solution to the DF acting on a circularly-moving perturber in a gaseous environment. Our solution exhibits a Coulomb divergence at short distances.

We begin with a critical look at the Green function approach to the computation of DF. Our treatment is entirely non-relativistic. 

\section{Method}

\subsection{Green function formulation}
\label{sec:Green}

\begin{figure}
  \includegraphics[width=9cm]{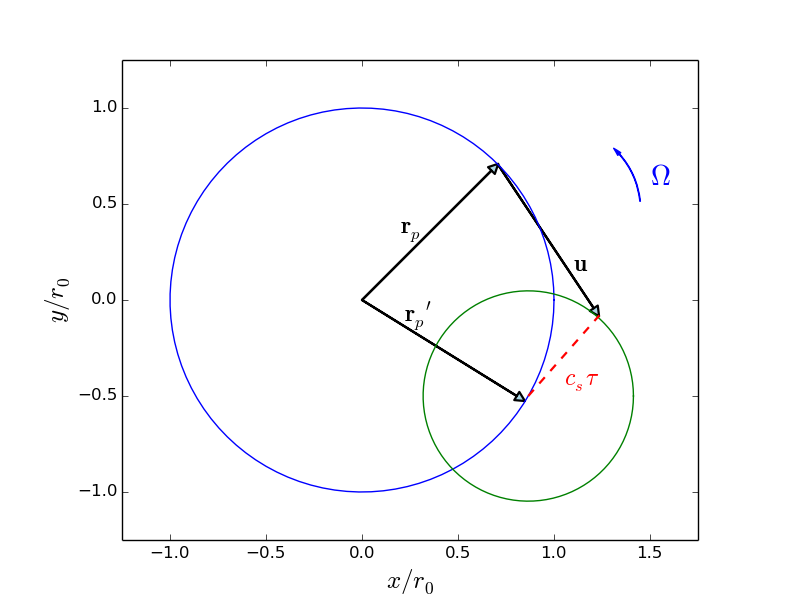} 
  \caption{The perturber rotates counter-clockwise with circular frequency $\Omega$ in the $x-y$ plane. At time $t'<t$, it was at position $\vr_p(t')\equiv {\vr_p}'$ and induced a sound wave that propagated over a distance $c_s\tau$ during the time interval $\tau\equiv t-t'$. The illustration assumes a Mach number $\mach>1$ and displays one possible $\vu$. The latter connects $\vr_p$ to any point on the green surface and, therefore, must be summed over as in Eq.~(\ref{eq:Generalfdf}).}
  \label{fig:setting}
\end{figure}

Consider an (infinite) uniform gaseous medium of density $\bar\rho_g$ and sound speed $c_s$.
The medium is perturbed by  a point mass in a fixed  circular orbit of radius $r_0$ and angular velocity $\Omega>0$, in the $x-y$ plane.
Let $\vr_p=\vr_p(t)$ and ${\vr_p}'=\vr_p(t-\tau)$ be the position of the perturber at "present-day" time $t$ and "retarded" time $t'=t-\tau$, respectively. 
 We have $\vr_p(t) = r_0(\cos\Omega t\,\xvh+\sin\Omega t\,\yvh)$. The Mach number is $\mach\equiv v_p/c_s$, where $v_p = \Omega r_0$ is the perturber's circular velocity.
Furthermore, let $\vu$ be the (three-dimensional) separation vector between $\vr_p(t)$ and the wavefront of the sound waves produced at time $t'$ (see Fig.~\ref{fig:setting} for a schematic illustration). 

%------------------------------------------------------------
%------------------------------------------------------------

Restricting the analysis to a linear response of the medium \cite[as in][]{ostriker:1999} leads to the driven sound wave equation 
\begin{equation}
    \frac{\partial^2\alpha}{\partial t^2} - c_s^2 \nabla^2\alpha = 4\pi G M \, h(t)\, \delta_D\!(\vr-\vr_p(t))
\end{equation}
where $\alpha(\vu,t)>0$ is the response fractional overdensity.
The Green function method can be applied in various ways to solve for $\alpha(\vu,t)$ and compute $\fdf(t)$. One possibility is to calculate $\alpha$
explicitly through the relation \citep{ostriker:1999}
\begin{align}
    \alpha(\vu,t) &= \frac{GM}{c_s^2}\int_{-\infty}^{+\infty}\!d\tau\, h(t-\tau) \frac{\dirac\!\big(\tau-\frac{1}{c_s}|\vr_p+\vu-{\vr_p}'|\big)}{|\vr_p+\vu-{\vr_p}'|}
    \nonumber \\
    &= \frac{GM}{c_s^2} \sum_{\tau_i>0} \frac{h(t-\tau_i)}{\big\lvert \vr_i - \frac{1}{c_s}\vr_i\cdot\vv_i\big\lvert} 
    \label{eq:GeneralWake} \;.
\end{align}
The function $h(t)$ determines when the perturber is on or off. In a {\it steady state} we have $h=1$ at all times  whereas, for a {\it finite time perturbation}, $h(t)=1$ for $t>0$ and zero otherwise. The retarded Green function $\frac{1}{r}\dirac(\tau-r/c_s)$ of the (sound) wave equation ensures causality, that is, the absence of solutions for $\tau < 0$. The second equality, in which $\vr_i\equiv\vr_p(t)+\vu-\vr_p(t-\tau_i)$ and $\vv_i\equiv\frac{d\vr_p}{dt}(t-\tau_i)$, expresses $\alpha(\vu,t)$ in terms of the Li\' enard-Wiechert (LW) potential \cite[e.g.][]{landau/lifshitz:1975}, a formulation convenient for numerical implementations. The roots $\tau_i$ are the time intervals for which the argument of the Dirac distribution vanishes.
The drag force then follows from  \begin{equation}
    \label{eq:Generalfdf}
    \fdf(t) = GM\bar\rho_g \int\!d^3u\,\frac{\vu}{u^3}\,\alpha(\vu,t) \;.
\end{equation}
Observe that $\fdf(t)$ is the gradient of a time-dependent potential so that $\oint\fdf\cdot d\vr\ne 0$.

\begin{figure}
  \includegraphics[width=9cm]{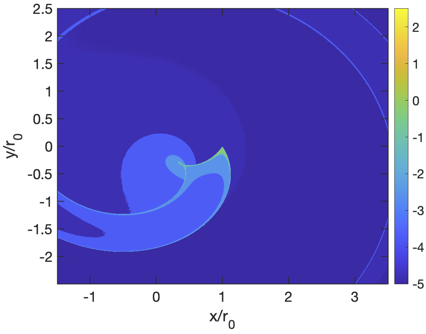}
  \caption{Density wake around a perturber moving circularly with Mach number $\mach=2$. Results show the overdensity $\alpha$ at time $t=4\pi/\Omega$ in the orbital plane for the finite time perturbation. $\alpha$ is singular on the Mach cone visible near the perturber. The latter is located at $\vr_p(t)=(1,0)$. Colored density contours indicate $\log\alpha$.}
  \label{fig:wake}
\end{figure}

In the supersonic case $\mach>1$, the LW potential is singular on the Mach cone near the perturber \citep[see Appendix A of][]{kim/etal:2007}. More precisely, if $s=r_0\Omega\tau$ is the path length along the orbit (measured from the current perturber's position), DF computed from the LW potential approximately scales as
\begin{equation}
    \fdf(t) \sim \int\!\frac{ds}{s}\Big(\sin s\, \rvh(t) + \cos s\,\pvh(t)\Big)
\end{equation}
in the vicinity of the perturber.
Here, $\rvh(t)$ is a unit vector in the  radial direction from the center of circular motion to the position of the perturber at $t$, 
and $\pvh(t)$ is a unit vector in the
tangential direction (parallel to the velocity). Clearly, a divergence will arise in the $\pvh$-direction as $s\to 0$.
This can be spotted in Fig.~\ref{fig:wake}, which shows the density wake in the orbital plane of a perturber moving circularly with Mach number $\mach=2$.
Therefore, we expect that a Green function implementation recovers the radial DF, but generally fails at predicting the tangential DF in numerical investigations.

To tackle this complication, we  consider a different route in which $\fdf(t)$ is evaluated directly. For this purpose, we express the retarded Green function as the Fourier transform
\cite[see][]{landau/lifshitz:1975,jackson:1975}
\begin{equation}
  \label{eq:FourierGreen}
    \frac{1}{r}\dirac\!\left(\tau-\frac{r}{c_s}\right)=4\pi c_s^2 \int_{\vk}\int_\omega \frac{e^{i(\vk\cdot\vr-\omega\tau)}}{c_s^2 k^2- \big(\omega+i\epsilon\big)^2}\;,
\end{equation}
where our convention is $\int_\omega \equiv \frac{1}{2\pi}\int_{-\infty}^{+\infty}\!d\omega$ and $\int_{\vk}=\frac{1}{(2\pi)^3}\int_0^{2\pi}\!d\varphi_k\int_{-1}^{+1}\!d\cos\vartheta_k\int_{-\infty}^{+\infty}\! dk k^2$ (in spherical coordinates $\vk=(k,\vartheta_k,\varphi_k)$). Moreover,  $\epsilon>0$ so that the resonant poles (i.e. resonances) are in the lower-half of the complex $\omega$ plane to ensure causality. The advantages of a Fourier treatment over a configuration space approach are i) to bypass the computation of possibly singular LW potentials and ii) to regularize the resonances by contour deformation.

As we shall see shortly, the divergence at $\mach>1$ remains, but it is encrypted in a maximum multipole $\ell_\text{max}$ that plays the role of small-scale cutoff.

\subsection{Helicity decomposition}
\label{sec:Helicity}

We start from Eq.~(\ref{eq:Generalfdf}). After substituting $\frac{1}{r}\dirac(\tau-r/c_s)$ by its Fourier space expression (\ref{eq:FourierGreen}), we replace both $e^{i\vk\cdot\vr_p}$ and $e^{-i\vk\cdot{\vr_p}'}$ by their respective Rayleigh expansions (see Appendix \S\ref{app:helicity} for details of the calculation). However, we retain $e^{i\vk\cdot\vu}$ and take advantage of the Fourier transform of the Coulomb potential to perform the integral over $\vu$:
\begin{equation}
    \int\!d^3u\,\frac{\vu}{u^3} e^{i\vk\cdot\vu} = 4\pi i \frac{\vk}{k^2} \;.
\end{equation}
It is convenient to decompose the wavemode $\vk$ onto the helicity basis $(\zvh,\ve_+,\ve_-)$ where $\ve_\pm=\frac{1}{\sqrt{2}}\big(i\yvh\mp\xvh\big)$ are the positive and negative helicity modes. Namely,
\begin{equation}
  \label{eq:helicityk}
  \vk = \sqrt{\frac{4\pi}{3}} k \left(Y_1^0(\kvh)\zvh+Y_1^{+1}(\kvh)\ve_+ +Y_1^{-1}(\kvh)\ve_-\right) \;,
\end{equation}
with $Y_l^m(\rvh)=Y_l^m(\vartheta,\varphi)$ being the usual spherical harmonics.
Likewise, the drag force decomposes into 
\begin{equation}
  \label{eq:helicityF}
  \fdf(t) = F^{(0)}\!(t)\,\zvh + F^{(+1)}\!(t)\,\ve_+ + F^{(-1)}\!(t)\,\ve_- \;.
\end{equation}
We must have $F^{(0)}\!(t) \equiv 0$ since the circular motion takes place in the plane $x-y$. Furthermore, we also have $F^{(+1)*}=-F^{(-1)}$ since the drag force is a real vector.

%------------------------------------------------------------
%------------------------------------------------------------

The Rayleigh expansions bring about factors of $Y_l^{m*}(\rvh_p)$ and $Y_l^{m*}({\rvh_p}')$. Since the azimuthal angle satisfies $\varphi_p - \varphi_p'=\Omega \tau$, we find $Y_l^{m*}({\rvh_p}')=e^{i m\Omega\tau} Y_l^{m*}(\rvh_p)$. On substituting this relation into the Rayleigh expansion of $e^{-i\vk\cdot{\vr_p}'}$ and performing the angle integral over $\kvh$ (which reduces to the Gaunt integral), we arrive at
\begin{align}
\label{eq:general}
F^{(m)}\!(t) &= 32\pi \left(\frac{GM}{c_s}\right)^2\bar\rho_g\sum_{l_1,m_1}\sum_{l_2,m_2}(-1)^{\frac{l_1-l_2+1}{2}} \nonumber \\
&\qquad \times \sqrt{(2l_1+1)(2l_2+1)}\,
Y_{l_1}^{m_1*}(\rvh_p) Y_{l_2}^{m_2*}(\rvh_p)
\nonumber \\ 
&\qquad \times
\left(\begin{array}{ccc} l_1 & l_2 & 1 \\ 0 & 0 & 0\end{array}\right) 
\left(\begin{array}{ccc} l_1 & l_2 & 1 \\ m_1 & m_2 & m\end{array}\right) 
S_{l_1,l_2}^{m_2} \;,
\end{align}
where the brackets denote Wigner 3j symbols, $j_l(x)$ are spherical Bessel functions, and
\begin{align}
    S_{l_1,l_2}^{m_2}&\equiv\lim_{\epsilon\to 0^+}\int_\omega \int_{-\infty}^{+\infty}\!d\tau\, h(t-\tau)e^{i(m_2\Omega-\omega)\tau} \nonumber \\
&\qquad \times 
c_s^2\int_0^\infty\!dk\,k\frac{j_{l_1}\!\big(kr_0\big)\, j_{l_2}\!\big(k r_0\big)}{c_s^2k^2-\big(\omega+i\epsilon\big)^2} 
\label{eq:scattering}
\end{align}
is the "scattering" amplitude of radial (standing) waves. Note that $S_{l_1,l_2}^{m_2*} = S_{l_1,l_2}^{-m_2}$.

The symmetry properties of the Wigner 3j symbols imply that there is a non-vanishing contribution only if $\ell_1=\ell_2\pm 1$ so that
$Y_{l_1}^{m_1*}(\rvh_p)Y_{l_2}^{m_2*}(\rvh_p)\propto e^{i m\Omega t}$. 
This reflects conservation of angular momentum, and guarantees that DF has the same frequency $\Omega$ as the circular motion. Furthermore, one can check that the sum 
\begin{equation}
    \sum_{m_1}\left(\begin{array}{ccc} l_1 & l_2 & 1 \\ m_1 & m_2 & m\end{array}\right) Y_{l_1}^{m_1*}(\rvh_p)Y_{l_2}^{m_2*}(\rvh_p) S_{l_1,l_2}^{m_2}
\end{equation}
vanishes identically for $m=0$ owing to the properties of the Wigner 3j symbols (they are non-zero only if $l_1+l_2+1$ is even) and to the fact that the polar angle is $\vartheta_p=\pi/2$. This ensures that the $z$-component of DF satisfies $F^{(0)}(t)\equiv 0$, as expected.

The components of DF in the helicity basis Eq.~(\ref{eq:helicityF}) can be recast into
\begin{align}
F^{(+1)}\!(t) &= 4\pi \left(\frac{GM}{\Omega r_0}\right)^2\bar\rho_g \frac{e^{i\Omega t}}{\sqrt{2}}I(\mach) \nonumber \\
F^{(-1)}\!(t) &= -F^{(+1)*}\!(t) \;.
\label{eq:helicityFIM}
\end{align}
In order to facilitate the comparison with \cite{ostriker:1999}, we have introduced the (generally complex) dimensionless function
\begin{align}
    I(\mach) &\equiv \frac{2\sqrt{2}}{\pi}\,\mach^2\sum_{l_1,m_1}\sum_{l_2,m_2} (-1)^{\frac{l_1-l_2+1}{2}} (2l_1+1)(2l_2+1)  \nonumber \\ & \qquad \times \sqrt{\frac{(l_1-m_1)!}{(l_1+m_1)!}\frac{(l_2-m_2)!}{(l_2+m_2)!}} \,
    P_{l_1}^{m_1}(0)\, P_{l_2}^{m_2}(0) \nonumber \\ & \qquad \times \left(\begin{array}{ccc} l_1 & l_2 & 1 \\ 0 & 0 & 0\end{array}\right)\left(\begin{array}{ccc} l_1 & l_2 & 1 \\ m_1 & m_2 & 1\end{array}\right) S_{l_1,l_2}^{m_2} \;.
\end{align}
Here, $P_l^m(x)$ are associated Legendre polynomials.

\subsection{General expression for circular motion}
\label{sec:GeneralResult}

Using the properties of the Wigner 3j symbols and the associated Legendre polynomials, the dimensionality of the multiple sum can be reduced, and the function $I(\mach)$ thereby simplified to 
\begin{widetext}
\begin{equation}
    \label{eq:IM}
    I(\mach) = \mach^2\sum_{l=1}^\infty \sum_{m=-l}^{l-2}(-1)^m\,\frac{(l-m)!}{(l-m-2)!}\,\frac{\Big(S_{l,l-1}^m - S_{l,l-1}^{m+1 *}\Big)}
    {\Gamma\!\left(\frac{1-l-m}{2}\right)\Gamma\!\left(1+\frac{l-m}{2}\right)\Gamma\!\left(\frac{3-l+m}{2}\right)\Gamma\!\left(1+\frac{l+m}{2}\right)} \;,
\end{equation}
\end{widetext}
which has much faster convergence than Eq.~(\ref{eq:general}). This expression is valid both in the subsonic and supersonic regime. 

The DF force eventually reads
\begin{equation}
  \label{eq:FDFIM}
    \fdf(t) = -4\pi \left(\frac{GM}{\Omega R}\right)^2\!\bar\rho_g\Big(\Re(I)\,\rvh(t)+\Im(I)\,\pvh(t)\Big)
\end{equation}
where $\Re(I)$ and $\Im(I)$ are the real and imaginary part of $I(\mach)$.
Eq.~(\ref{eq:FDFIM}), together with $I(\mach)$ given by Eq.~(\ref{eq:IM}) and $S_{\ell,\ell-1}^m$ given by Eq.~(\ref{eq:scattering}), is the central result of this paper.

\subsection{Coulomb divergence}
\label{sec:CoulombDivergence}

The analyticity of the Green function along with the Sokhatsky-Weierstrass theorem
\begin{equation}
    \lim_{\epsilon\to 0^+}\frac{1}{x\pm i\epsilon} = P \frac{1}{x} \mp i\pi\dirac(x)\;,
\end{equation}
where $P$ denotes the Cauchy principal value, implies a connection between $\Re(I)$ and $\Im(I)$ or, equivalently, between the radial and tangential components of DF which is yet another manifestation of the Kramers-Kronig relations. However, while this implies that the real and imaginary part of the scattering amplitudes determine each other, the infinite sum (\ref{eq:IM}) does not guarantee that both $\Re(I)$ and $\Im(I)$ are finite.

\begin{figure}
  \centering
  \includegraphics[width=9cm]{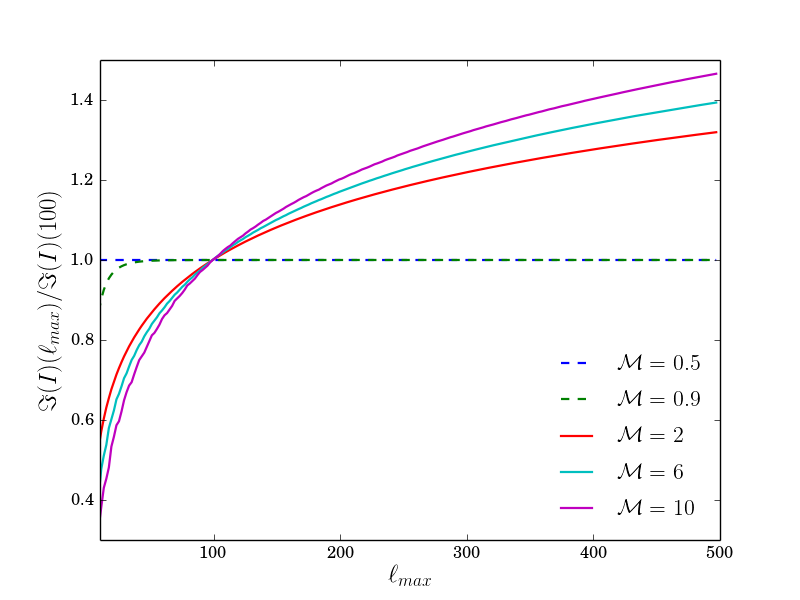} 
  \caption{The tangential DF component in the steady-state regime is shown as a function of the maximum multipole $\ell_\text{max}$ for a few values of $\mach$. The different curves are all normalized to unity at $\ell_\text{max}=100$. A logarithmic divergence is present in the supersonic case. It increases mildly with Mach number.}
  \label{fig:Convergence}
\end{figure}

To illustrate this point, Fig.~\ref{fig:Convergence} displays $\Im(I)$ as a function of the maximum multipole $\ell_\text{max}$ at which the sum (\ref{eq:IM}) is truncated, with $\ell_\text{max}$ as large as 500. Results are shown for a few values of $\mach$. They demonstrate that $\Im(I)$ has a logarithmic divergence in the supersonic regime, in accordance with the numerical results of \cite{kim/etal:2007} and the recent analytical study of \cite{fouvry/etal:2021}. The magnitude of this divergence increases mildly with $\mach$. In addition, we have numerically checked that $\Re(I)$ (not shown in Fig.~\ref{fig:Convergence}) converges regardless of the value of $\mach$.

%------------------------------------------------------------
%------------------------------------------------------------

This shows that our approach recovers the small-scale, Coulomb divergence rooted in the singular nature of the LW potentials in the supersonic regime, in agreement with the linear result \citep{ostriker:1999}. The Fourier approach considered here has not removed this divergence, but the latter is now all encoded in the dependence of $I(\mach)$ on $\ell_\text{max}$. This will be helpful to assess the dependence of numerical simulations on resolution.

\section{Applications}

Our approach has reduced the problem to the evaluation of a scattering amplitude which, for a given Green function, depends on the choice of boundary/initial conditions.

\subsection{A single perturber in a circular motion}

In steady-state, the integral over the time shift $\tau$ in Eq.~(\ref{eq:scattering}) returns $(2\pi)\dirac(m_2\Omega-\omega)$. This leaves us with a scattering amplitude $S_{l,l-1}^{m_2}(\mach)\equiv S_{l,l-1}^{\text{sty}}(m_2\mach)$ given by
\begin{align}
    \label{eq:amplitudegassteady}
    S_{l,l-1}^\text{sty}\!(m_2\mach) &= \lim_{\epsilon\to 0^+}\int_0^\infty \! dz\,\frac{z\,j_l\!\big(z\big)\, j_{l-1}\!\big(z\big)}{z^2-\big(m_2\mach+i\epsilon\big)^2} \nonumber \\
    &= \frac{i\pi}{2}\, j_l(m_2\mach)\, h_{l-1}^{(1)}(m_2\mach)
\end{align}
where $z=kr_0$.
To perform the integral, we expressed the spherical Bessel functions in terms of the Hankel functions $h_l^{(1)}\!(z)$ and $h_l^{(2)}\!(z)$, i.e.  $j_l(z)=\frac{1}{2}(h_l^{(1)}\!(z)+h_l^{(2)}\!(z))$, and applied the method of Residues. Taking into account the contribution of a pole at $z=0$ to the contour integral of $h_{l_1}^{(1)}h_{l_2}^{(1)}$ (or $h_{l_1}^{(2)}h_{l_2}^{(2)}$), the resonant poles and the Wronskian $j_{l-1}(z)y_l(z)-j_l(z)y_{l-1}(z) = -\frac{1}{z^2}$ (here $y_l(z)$ is a modified spherical Bessel function) eventually lead to Eq.~(\ref{eq:amplitudegassteady}).
When $m_2\mach=0$, the scattering amplitude is solely controlled by the pole at the origin. In this case, taking the argument in Eq.~(\ref{eq:amplitudegassteady}) to zero yields
\begin{equation}
    S_{l,l-1}^{\text{sty}}\!(0) = \frac{\pi}{2(4l^2-1)}\label{eq:limitingS0} \;,
\end{equation}
which agrees with a direct Residues calculation, as expected.

The scattering amplitude $S_{l,l-1}^\text{sty}\!(m_2\mach)$ converges to zero in the limit of large arguments. This happens so long as the azimuthal number $m_2$ is different from zero. When $m_2=0$, the argument is zero for any finite value of $\mach$. As a result, only the terms with $m=0$ (from $S_{l,l-1}^m$) or $m=-1$ (from $-S_{l,l-1}^{m+1 *}$) in the series expansion Eq.~(\ref{eq:IM}) survive when $\mach\to\infty$. Since these are all proportional to $S_{l,l-1}(0)$, and since the steady-state amplitude $S_{l,l-1}^\text{sty}\!(0)$ is real and independent of $\mach$, we must have $\Re(I)\to {\rm const}\,\mach^2$ and $\Im(I)\to 0$ in the limit $\mach\to\infty$.

\begin{figure}
  \centering
  \includegraphics[width=9cm]{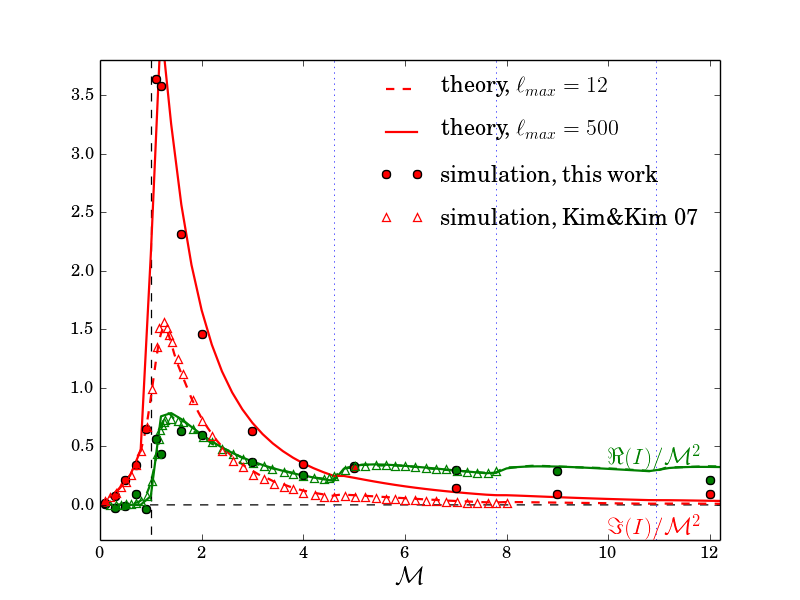} 
  \caption{The real (magenta) and imaginary (red) part $\Re(I)$ and $\Im(I)$ of Eq.~(\ref{eq:IM}) as a function of Mach number $\mach$ in the steady-state regime. The (black) dashed vertical lines indicates $\mach=1$. The (blue) dotted vertical lines mark the critical Mach numbers $\mach_n$ ($n=1,2,\dots$) at which the number of roots of the equation $\tau-r/c_s=0$ changes \citep[see][for the definition of $\mach_n$]{kim/etal:2007}. The series expansion (\ref{eq:IM}) is truncated at multipole $\ell_\text{max}$ as indicated on the Figure. The symbols show DF estimates extracted from two different numerical simulations with resolution $r_\text{min}/r_0=0.1$ \citep{kim/etal:2007} and $r_\text{min}/r_0\simeq 7\times 10^{-3}$ (our numerical runs, see Appendix \S\ref{app:simulation}).}
  \label{fig:IM}
\end{figure}

The real and imaginary parts of $I(\mach)$ are displayed in Fig.~\ref{fig:IM}. Our theoretical prediction (solid curves) is compared to the DF extracted from numerical schemes implementing the LW-based approach discussed in \S\ref{sec:Green}: triangles is data from \cite{kim/etal:2007}, circles indicate our own simulation results (see Appendix \S\ref{app:simulation} for details). For $\Re(I)$, the agreement between theory and simulations is excellent, which demonstrates that the latter have converged.
For $\Im(I)$ however, the numerical data depends on the actual resolution of the simulations which the Coulomb divergence is sensitive to. Truncating the series expansion (\ref{eq:IM}) at multipole $\ell_\text{max}=12$ and 500 yields a reasonable fit to the two sets of simulations. The inferred values of $\ell_\text{max}$ are broadly consistent with the naive scaling $\ell_\text{max} \sim \pi/(r_\text{min}/r_0)$, where the resolution is $r_\text{min}/r_0=0.1$ and $\simeq 7\times 10^{-3}$ in \cite{kim/etal:2007} and in our simulations, respectively.

%------------------------------------------------------------
%------------------------------------------------------------

While there is a short-distance (ultraviolet) divergence in the supersonic case, there is no long-distance (infrared) divergence at all in the circular case.
To understand this, observe that the denominator of Eq.~(\ref{eq:amplitudegassteady}) saturates for $z\lesssim m_2\mach$, i.e. $k\lesssim m_2/c_sT$ where $T$ is the orbital period, so that the integrand always vanishes in the limit $z\to 0$. In plain words, long-wavelength fluctuations do not contribute to the DF because they appear homogeneous to the circular perturber (the density pattern repeats itself on scales $\gtrsim c_s T$). This removes any long-distance (infrared) divergence.

Turning to the finite time perturbation, we have $h(t-\tau)=0$ for $\tau > t$. Consequently, the integral over the variable $\tau$ reduces to
\begin{equation}
    \int_{-\infty}^t\!d\tau\,e^{i(m_2\Omega-\omega)\tau} = \lim_{\eta\to 0^+}
    \frac{e^{i(m_2\Omega-\omega)t}}{i(m_2\Omega-\omega-i\eta)}\;.
\end{equation}
As a result, the scattering amplitude is the sum $S_{l,l-1}^{\text{ftp},m_2} = S_{l,l-1}^{\text{sty},m_2}(\mach)+S_{l,l-1}^{\text{tra},m_2}\!(\mach,t)$, in which the transient contribution is the limit $\epsilon\to 0^+$ of
\begin{align}
 S_{l,l-1}^{\text{tra},m_2}\!(\mach,t)= 
    -\frac{e^{2 i m_2\mach \tilde t}}{2}
    \int_{-\infty}^{+\infty}\!dz\,e^{2iz \tilde t}
    \frac{j_l(z)\, j_{l-1}(z)}{k+m_2\mach+i\epsilon} 
\end{align}
Here, $\tilde t = t/t_\text{sc}$ is in unit of the sound crossing time $t_\text{sc}=2r_0/c_s$.
$S_{l,l-1}^{\text{tra},m_2}\!(\mach,t)$ can also be directly evaluated using the method of Residues. The decomposition of the integrand into a sum of products of Hankel functions reveals that the transient contribution vanishes for $\tilde t>1$, in agreement with the numerical findings of \cite{kim/etal:2007}.
This explains why, in the limit $\mach\ll 1$, the steady-state circular solution $\Im(I)\approx \frac{1}{3}\mach^3$ recovers the drag created by a linear trajectory perturber turned on at $t=0$: for $t_\text{sc}< t \ll t_\text{sc}/\mach$, steady-state is achieved and, at the same time, $t$ is small enough that the motion is approximately linear. 

Obtaining a compact expression for $S_{l,l-1}^{\text{tra},m_2}\!(\mach,t)$ is, however, challenging owing to the pole of order $2l+1$ at the origin.

%------------------------------------------------------------
\begin{figure}
  \centering
  \includegraphics[width=9cm]{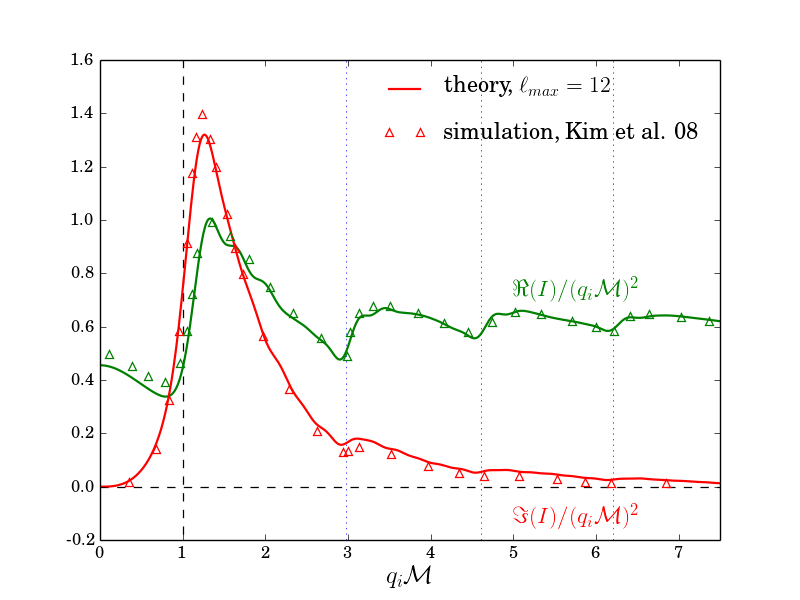} 
  \caption{The tangential (red) and radial (magenta) DF components in the steady-state regime for a compact circular binary with equal mass, $q_1=q_2=1/2$ (see text). Multipoles up to $\ell_\text{max}=12$ are included in the evaluation of $I(\mach)$ in order to match the resolution of the simulation results shown here \citep[the data points were extracted from][]{kim/etal:2008}.}
  \label{fig:Binary}
\end{figure}
%------------------------------------------------------------

\subsection{Compact circular binary}

The previous results straightforwardly extend to a compact binary on a circular orbit. We only need to take into account the presence of a second source at the antipode of $\vr_p'$, that is, with an azimuthal angle $\varphi'+\pi = \Omega(t-\tau)+\pi$. Let $M$ be the total binary mass and $q_1M$, $q_2M$ (with $q_1+q_2=1$) be the  masses of the compact objects. Since the latter generally differ, so do their distances from the binary center-of-mass (assumed to be at rest relative to the gaseous medium). This affects the argument of the spherical Bessel functions in the Rayleigh expansion. For this purpose, we introduce the generalized amplitude
\begin{equation}
    \label{eq:extendedamplitude}
    S_{l,l-1}^{\alpha,\beta}\!(x)\equiv \lim_{\epsilon\to 0^+}\int_0^\infty\!dz\,\frac{z\,j_l\!\big(\alpha z\big)\, j_{l-1}\!\big(\beta z\big)}{z^2-\big(x+i\epsilon\big)^2}
\end{equation}
whose magnitude depends on whether $\alpha>\beta$ or $\alpha<\beta$. We find
\begin{equation}
    \label{eq:extendedamplituderes1}
    S_{l,l-1}^{\alpha,\beta}\!(x) =\left\lbrace\begin{array}{cc}
    \frac{\pi}{2}\,\bigg[i\, j_{l-1}\!\big(\beta x\big)\, h_l^{(1)}\!\big(\alpha x\big)-\left(\frac{\beta^{l-1}}{\alpha^{l+1}}\right)\frac{1}{x^2}\bigg] & (\alpha>\beta) \\
    \frac{i\pi}{2}j_l(\alpha x) h_{l-1}^{(1)}\!(\beta x) & (\alpha < \beta)
    \end{array}\right.
\end{equation}
and, in particular,
\begin{equation}
    S_{l,l-1}^{\alpha,\beta}\!(0) = \left\lbrace\begin{array}{cc}\frac{\pi}{4}\left(\frac{\beta^{l-1}}{\alpha^{l+1}}\right)
    \frac{\big[(1+2l)\alpha^2+(1-2l)\beta^2\big]}{(4l^2-1)} & (\alpha > \beta) \\
    \frac{\pi}{2}\left(\frac{\alpha}{\beta}\right)^l \frac{1}{(4l^2-1)} & 
    (\alpha <\beta) \end{array}\right.
\end{equation}
The computation proceeds along the lines of $S_{l,l-1}^\text{sty}\!(x)$.
Exploiting a Wronskian relation of Bessel functions consistently returns $S_{l,l-1}^{\alpha,\alpha}\!(x)=\frac{i\pi}{2}j_l(\alpha x)h_l^{(1)}\!(\alpha x)$ in the special case $\alpha=\beta$.

We define the Mach number of the binary system as $\mach = \Omega r_0/c_s$, where $r_0$ is the circular radius. The function $I(\mach)$ can be recast in the form Eq.~(\ref{eq:IM}) provided that the difference $S_{l,l-1}^m-S_{l,l-1}^{(m+1)*}$ in the numerator is replaced by
\begin{multline}
    \beta^2\Big[S_{l,l-1}^{\alpha,\alpha}\!\big(m\mach\big)-S_{l,l-1}^{\alpha,\alpha*}\!\big((m+1)\mach\big)\Big] \\ +(-1)^m\alpha\beta\Big[S_{l,l-1}^{\alpha,\beta}\!\big(m\mach\big)+S_{l,l-1}^{\alpha,\beta*}\!\big((m+1)\mach\big)\Big]
\end{multline}
with $(\alpha,\beta)=(q_2,q_1)$ or $(q_1,q_2)$ depending on whether $I(\mach)$ encodes the DF acting on the compact object of mass $q_1M$ or its companion, respectively. The components of the DF can still be cast in the form of Eq.~(\ref{eq:helicityFIM}), except that $M$ now is the total mass of the compact binary. Like in the single perturber case, the tangential DF for a compact circular binary suffers from a small-scale logarithmic divergence. 

To facilitate the comparison with \cite{kim/etal:2008}, who extended the approach of \cite{kim/etal:2007} to a compact binary, Fig.~\ref{fig:Binary} displays our prediction for the equal mass case $q_1=q_2=1/2$ and assuming $\ell_\text{max}=12$ (the simulations of \cite{kim/etal:2008} have the same resolution as those of \cite{kim/etal:2007}). This low value of $\ell_\text{max}$ is responsible for the wiggles seen in the theoretical curves. Overall, our prediction closely track the smooth shape and the features seen in the numerical data of \cite{kim/etal:2008}. 

\section{Conclusions}

We have shown that the dynamical friction acting on circularly moving perturbers can be cast in a compact form easily amenable to numerical evaluations. Although we have focused on individual point mass  perturbers and compact circular binaries in a gaseous environment, our analytical method is not restricted to these specific cases. It provide a versatile tool for exploring dynamical friction in different dynamical systems and environment so long as the typical size $R$ of the system satisfies $R \gtrsim GM/c_s^2$ (to ensure the validity of linear response theory). However, one should bear in mind that nonlinearities, which are neglected in our current approach, may result in a DF different from that obtained here \cite[see, e.g., ][for a nonlinear treatment of DF acting on an accreting body]{lee/stahler:2011}.

Our analytical approach provides insights into the nature of dynamical friction for circular motions. In particular, it recovers the Coulomb (logarithmic) divergence for supersonic motion in steady-state and finite time perturbation regime, in accordance with three-dimensional simulations implementing the Li\'enard-Wiechert potentials. However, we expect that the presence of a small-scale divergence critically depends on the $k$-dependence of the Green function at large wavenumbers. Therefore, what is true in the gaseous case may not hold for other media.
Our analysis also shows that, when the perturber is turned on at $t=0$ (finite time perturbation case), steady-state is achieved after one sound-crossing time. This suggests that our steady-state solution should have interesting astrophysical applications. It could be relevant for the DF on galaxies moving near the gaseous cores of clusters of galaxies, or compact binaries in gaseous environments for instance.

We thank anonymous referees for helpful reports.
This work was supported in part by the Israel Science Foundation (ISF) grants no 2562/20 (VD and RB) and 936/18 (AN).

\vspace{2mm}

{\bf Data availability}

\vspace{2mm}

The data that support the findings of this study are openly available at the following URL: \url{https://github.com/nyalothep/dynfricGas} .

\appendix

\section{Helicity decomposition of DF}
\label{app:helicity}

We begin with the substitution of the Fourier space Green function in Eq.~(\ref{eq:GeneralWake}), set $\vr=\vr_p+\vu-\vr_p'$ and write Eq.~(\ref{eq:Generalfdf}) as
\begin{align}
  \fdf(t) &= 4\pi (G M)^2 \rho_g \int_\omega \int_{-\infty}^{+\infty}\! d\tau\,h(t-\tau) \int_{\vk}
  \frac{e^{i\vk\cdot(\vr_p-\vr_p')}}{c_s^2 k^2 - (\omega+i\epsilon)^2} \int\!d^3u\,\frac{\vu}{u^3}\, e^{i\vk\cdot\vu}
  \nonumber \\
  &= 4\pi (G M)^2 \rho_g \int_\omega \int_{-\infty}^{+\infty}\! d\tau\,h(t-\tau) \int_{\vk}
  \frac{e^{i\vk\cdot(\vr_p-\vr_p')}}{c_s^2 k^2 - (\omega+i\epsilon)^2} \left(4\pi i \frac{\vk}{k^2}\right) \;.
\end{align}
Next, we expand both $e^{i\vk\cdot\vr_p}$ and $e^{-i\vk\cdot\vr_p'}$ in plane waves according to 
\begin{equation}
  e^{i\vk\cdot\vr} = 4\pi \sum_{\ell=0}^\infty\sum_{m=-\ell}^{+\ell} i^\ell j_\ell(kr) Y_\ell^m(\kvh) Y_\ell^{m*}(\rvh)
\end{equation}
and decompose the vector $\vk$ onto the helicity basis $(\zvh,\ve_+,\ve_-)$ as in Eq.~(\ref{eq:helicityk}), where $\ve_\pm = \frac{1}{\sqrt{2}}(i\yvh\mp\xvh)$ are
complex vectors that lie in the orbital plane $x - y$. Inserting these relations into the expression of $\fdf(t)$ and splitting the integral over $\vk$ into a radial
($k$) and an angular ($\kvh$) part yields
\begin{align}
  \fdf(t) &= (4\pi)^4 \big(GM\big)^2 \rho_g \sum_{\ell_1 m_1} \sum_{\ell_2 m_2} i^{\ell_1+1} (-i)^{\ell_2} \int_{-\infty}^{+\infty}\!\frac{d\omega}{2\pi} \int_{-\infty}^{+\infty}\!d\tau\, h(t-\tau) e^{-i\omega\tau}
  \int_0^\infty\!\frac{dk}{2\pi^2}\,\frac{k j_{\ell_1}(k r_0) j_{\ell_2}(k r_0)}{c_s^2 k^2 - (\omega+i\epsilon)^2} \nonumber \\
  &\qquad \times \int \!\frac{d^2\kvh}{4\pi}\,\sqrt{\frac{4\pi}{3}}\left(Y_1^0(\kvh)\zvh+Y_1^{+1}(\kvh)\ve_+ + Y_1^{-1}(\kvh)\ve_-\right)
  Y_{\ell_1}^{m_1}(\kvh) Y_{\ell_1}^{m_1 *}(\rvh_p) Y_{\ell_2}^{m_2}(\kvh) Y_{\ell_2}^{m_2 *}(\rvh_p') \label{eq:fdftmp1} \;,
\end{align}
where $\rvh_p$ and $\rvh_p'$ are the unit vectors aligned with $\vr_p$ and $\vr_p'$, respectively,
and $\int\!d^2\kvh = \int_0^{2\pi}\!d\varphi_k\int_{-1}^{+1}\!d\cos\vartheta_k$.
In spherical coordinates $(r,\vartheta,\varphi)$,
$\vr_p=(r_0,\pi/2,\Omega t)$ and $\vr_p'=(r_0,\pi/2,\Omega (t-\tau))$ so that $Y_{\ell_2}^{m_2 *}(\rvh_p')= e^{i m_2 \Omega\tau} Y_{\ell_2}^{m_2 *}(\rvh_p)$.
On inserting this relation into Eq.~(\ref{eq:fdftmp1}) and carrying out the angular integration over $\kvh$ with aid of the Gaunt integral,
we can read off Eq.~(\ref{eq:helicityF}) with helicity components $F^{(m)}(t)$ given by Eq.~(\ref{eq:general}).

\section{Numerical Scheme }
\label{app:simulation}

The fractional density perturbation $\alpha(\mathbf{x},t) =\rho_g(\mathbf{x},t)/\bar \rho_g-1 $ obeys the linear wave equation \citep[e.g.,][]{ostriker:1999}
\begin{equation}
    c_s^2\nabla^2\alpha-\frac{\partial^2\alpha}{\partial t^2}=-4\pi \rho_\text{ext}
\end{equation}
where $\rho_\text{ext}(t,\vr)=M \delta_D\!(\vr-\vr_p(t)) h(t)$ is the external density perturbation. 
For a point mass perturbation with Mach number $\mach<1$, the solution is given by~\cite{landau/lifshitz:1975}
\begin{equation}
\label{eq:solsingle}
    \alpha(\vx,y)=\frac{GM}{c_s^2}\left(r-\frac{\mathbf{v}\cdot\vr}{c_s}\right)^{-1} \;.
\end{equation}
Here, $\vr=\vx-\vr_p(t')$ and $r=|\vr|$, where $\vr_p(t') $  and  $\vv=\dot{\vr}_p(t')$ are the location and  velocity of the perturber at time $t'$. The retarded time $t'$ satisfies the equation
$t'+r(t')/c_s=t$. For subsonic motion, there is only one root to this equation, while for the supersonic motion it is satisfied by multiple solutions. 
In this case the density is obtained as the sum  over all roots of the expression on the r.h.s in \ref{eq:solsingle}.
Fig.~\ref{fig:wake} displays the resulting density wake in the orbital plane of a perturber moving circularly with Mach number $\mach=2$. 

\cite{kim/etal:2007,kim/etal:2008} compute the density in a three-dimensional volume around the perturber by numerically solving for the roots in the finite perturbation time case. They then compute the gravitational force affected by the density field on the perturber. 
Here, we follow a different approach which mitigates the intensive {\small CPU} and memory requirement of the problem. We divide the orbital plane into a two-dimensional Cartesian ($x-y$) grid of resolution $r_\text{min}$. The equation for $t'$ is recast as 
\begin{equation}
    z^2=c_s^2(t-t')^2-(x-x_p')^2-(y-y_p')^2\;  
\end{equation}
where $\vr_p(t')=(x_p',y_p',0)$. Instead of finding multiple solution for $t'$ given the multiplet $(t,x,y,z)$, we divide $t'$ into equal time steps, $t_n'$,  and find the corresponding $z=z(t_n',x,y)$ using the above equation. 
This avoids the complication related to locating multiple roots. 
Furthermore, in the calculation of the force, the integration over the $z$ coordinate is replaced by a summation over time taking into account that $dz=(dz/dt') dt'$.
This eliminates the need for large three-dimensional grids since the integration over $z$ in this manner can be done by adding up contributions to the density and the force at each time step individually.
In Fig.~\ref{fig:IM}, filled symbols indicate our results obtained for a $x-y$ grid resolution $r_\text{min}=5 r_0/700\simeq 7\times 10^{-3} r_0$.

\bibliography{references}

\end{document}